\documentclass[prd,twocolumn,nofootinbib,floatfix,preprintnumbers,letterpaper]{revtex4}
\usepackage{bm} 
\usepackage{graphicx,epsfig}
\usepackage{latexsym,amssymb,amsmath,float,url}
\usepackage{latexsym}
\newcommand\bea{\begin{eqnarray}}
\newcommand\eea{\end{eqnarray}}
\newcommand\aea{&=&}

\newcommand\psm{\phi_{SM}}
\newcommand\shs{\sigma_{H}}

\begin{document}

\title{ A Higgs--Saw Mechanism as a Source for Dark Energy}

\author{Lawrence M. Krauss}
\email{krauss@asu.edu}
\affiliation{School of Earth and Space Exploration \& Department of Physics, Arizona State University, Tempe, AZ, USA 85287-1404}
\affiliation{Research School of Astronomy and Astrophysics, Mt. Stromlo Observatory, Australian National University, Canberra, Australia 2614}

\author{James B.\ Dent}
\email{jbdent@louisiana.edu}
\affiliation{Department of Physics, University of Louisiana at Lafayette, Lafayette, LA 70504-4210, USA}

\begin{abstract}
Motivated by the see-saw mechanism for neutrinos which naturally generates small neutrino masses, we explore how a small GUT-scale mixing between the Standard Model Higgs and an otherwise massless hidden sector scalar can naturally generate a small mass and VEV for the new scalar which produces a false vacuum energy density contribution comparable to that of the observed dark energy dominating the current expansion of the Universe.  The new scalar produces no discernible signatures in existing terrestrial experiments so that one may have to rely on other cosmological tests of this idea.
\end{abstract} 

\pacs{xxxx-x}


\maketitle 

\section{Introduction}

One of the biggest challenges in particle physics today is to understand how new physics, presumably associated with energy scales that have not yet been directly probed at accelerators, can nevertheless generate a vacuum energy density which is as incredibly small as that required to produce Dark Energy to drive the current accelerated expansion of the Universe. 
  
We reconsider this problem in the context of another example where extremely small mass scales can be naturally generated in particle physics, namely the See-Saw Mechanism for neutrinos \cite{Minkowski:1977sc,Mohapatra:1979ia,Yanagida:1979aa,Glashow:1980aa,GellMann:1979aa,Foot:1988aq,Schechter:1980gr,Lazarides:1980nt}, which generates phenomenologically acceptable small neutrino masses through small mixing between otherwise decoupled neutrino eigenstates associated with physics at vastly different mass scales. 
 
Our launching point is the existence of a Higgs particle in the Standard Model, now verified at the LHC, which opens up the possibility that other fundamental scalars exist in nature, some of which might naturally mix with the Higgs.  There has been a tremendous amount of work done in exploring possible implications of a scalar sector which couples to the Higgs boson via what is known as the Higgs portal. (for early proposals see \cite{Patt:2006fw,Silveira:1985rk,Foot:1991bp})  Included in the literature are proposals ranging from generating the EW scale from a much higher scale \cite{Englert:2013gz,Calmet:2006hs,Heikinheimo:2013xua}, to coupling the Higgs to a Quintessence dark energy field \cite{Bertolami:2007wb,Bento:2009xa}, and most recently to possible cosmological consequences of new Goldstone bosons \cite{Weinberg:2013kea}. 
Here we take another complementary tack.  We consider whether a tiny dark energy scale can be generated from the EW scale by Higgs mixing with a field whose interactions are suppressed by a large energy scale, just as small neutrino masses are generated by the conventional see-saw mechanism.  We  demonstrate here that if massless scalars exist in sectors that are decoupled from the visible sector by GUT-scale suppressions, then a natural mechanism exists to generate masses and vacuum energy contributions precisely in the range of that currently associated with Dark Energy in the Universe.    This mechanism appears quite generic and simply requires a small mixing, suppressed by the ratio of the weak scale to the GUT scale between the Higgs ,and any otherwise massless scalar in a hidden sector.   

It is worth noting that this mechanism provides a natural way to produce a { \it contribution} to the vacuum energy that is of the correct magnitude of the observed dark energy. It does not however resolve the deeper longstanding problem of why the total vacuum energy, including contributions from  the visible sector, is not much larger.  In particular, the vacuum energy contribution that results from symmetry breaking is generically negative, so that if one were to resolve this problem by requiring the ultimate true vacuum to have zero energy, a positive vacuum energy density would require a false vacuum state today.   Note also that we do not discuss possible additional detailed physics of the hidden sector, including issues associated with massless scalars, possible radiative corrections, dark matter etc.   The simple model we provide is a proof of principle that a contribution corresponding to the observed magnitude of dark energy may arise naturally if additional scalar fields are mixed with the standard model Higgs with GUT-scale suppressions.

\section{Higgs Mixing: A Toy Model}

Let us assume for simplicity two scalar fields: $\phi_{SM}$ and $\sigma_{H}$ where $\phi_{SM}$ is charged under the Standard Model (SM) and $\sigma_{H}$ is charged under some hidden sector (H).  A simple form for the scalar potential for these can be written as (where $x^2$  is taken to imply $x^{\dagger}x$)
\bea
V(\phi_{SM},\sigma_{H}) = V(\psm) + \frac{\lambda_{mix}}{4}\phi_{SM}^2\sigma_{H}^2 +  \frac{\lambda_{H}}{4}\shs^4
\eea
where one can see that the two sectors are coupled with strength $\lambda_{mix}$.

Now one can go from the gauge eigenstates $\psm$ and $\shs$ to mass eigenstates, which we will call $h$ for the Higgs and $s$ for the other scalar.  The gauge eigenstates will take on vevs of
\bea
\langle\psm\rangle = \frac{v}{\sqrt{2}}\\
\langle\shs\rangle = \frac{v_{H}}{\sqrt{2}}
\eea
The gauge states have a mass matrix
\bea\label{gaugematrix}
\left(\begin{array}{cc}
m_{\phi}^2 & m_{\phi\sigma}^2\\
m_{\phi\sigma}^2 & m_{\sigma}^2
\end{array}
\right)
\eea
where
\bea
m_{\phi}^2 \aea \frac{\partial^2V}{\partial\psm^2}\bigg|_{\psm = \langle\psm\rangle\,;\shs = \langle\shs\rangle}\\
m_{\sigma}^2 \aea \frac{\partial^2V}{\partial\shs^2}\bigg|_{\psm = \langle\psm\rangle\,;\shs = \langle\shs\rangle}\\
m_{\phi\sigma}^2 \aea \frac{\partial^2V}{\partial\psm\partial\shs}\bigg|_{\psm = \langle\psm\rangle\,;\shs = \langle\shs\rangle}
\eea
One can rotate to the mass basis via
\bea
h \aea \psm\,\textrm{cos}\,\theta + \shs\,\textrm{sin}\,\theta\\
s \aea -\psm\,\textrm{sin}\,\theta + \shs\,\textrm{cos}\,\theta
\eea
where $\theta$ is the mixing angle.  The mass basis obviously has the diagonal mass matrix
\bea
\left(\begin{array}{cc}
m_{h}^2 & 0\\
0 & m_{s}^2
\end{array}
\right)
\eea
with eigenvalues arising from Eq.(\ref{gaugematrix}) given by
\bea
m_h^2 \aea \frac{m_{\phi}^2 + m_{\sigma}^2}{2} + \frac{m_{\phi\sigma}^2}{\epsilon}\sqrt{1+\epsilon^2}\\
m_s^2 \aea \frac{m_{\phi}^2 + m_{\sigma}^2}{2} - \frac{m_{\phi\sigma}^2}{\epsilon}\sqrt{1+\epsilon^2}
\eea
where
\bea
\epsilon \equiv \frac{2m_{\phi\sigma}^2}{m_{\phi}^2-m_{\sigma}^2}
\eea
One can then solve for the mixing angle by equating
\bea
\nonumber \left(\begin{array}{cc}
\psm & \shs
\end{array}\right)
\left(\begin{array}{cc}
m_{\phi}^2 & m_{\phi\sigma}^2\\
m_{\phi\sigma}^2 & m_{\sigma}^2
\end{array}
\right)
\left(
\begin{array}{c}
\psm\\
\shs
\end{array}
\right)  \\
 = 
\left(\begin{array}{cc}
h & s
\end{array}\right)
\left(\begin{array}{cc}
m_{h}^2 & 0\\
0 & m_{s}^2
\end{array}
\right)
\left(
\begin{array}{c}
h\\
s
\end{array}
\right) 
\eea
and one finds
\bea
\theta = \frac{1}{2}\textrm{tan}^{-1}\left(\frac{2m_{\phi\sigma}^2}{m_{\phi}^2-m_{\sigma}^2}\right) = \frac{1}{2}\textrm{tan}^{-1}\epsilon
\eea
A small mixing angle is obviously the same as small coupling $\lambda_{mix}$.

Minimizing the potential when the fields are set equal to their VEVs gives
\bea
v_{HS}\left(\frac{\lambda_{mix}}{2}v^2 + \lambda_{H}v_{H}^2\right) = 0
\eea
or
\bea
v_{H}^2 = -v^2\frac{\lambda_{mix}}{2\lambda_{H}}
\eea

A positive VEV requires $\lambda_{mix} < 0$ if we assume positivity of $\lambda_{H}$ so the $\sigma_H$ potential is bounded from below. 

One thus naturally gets a small hidden sector VEV  compared to the Standard Model VEV via small mixing, $\lambda_{mix} \ll \lambda_{H}$.   In the limit, $\epsilon \ll 1$ and from Eq (12) we find

\bea\label{smass1}
m_s^2= m_{\sigma}^2 -\frac{m_{\phi}^2}{4}\epsilon^2 + \frac{m_{\sigma}^2}{4}\epsilon^2 + \cdots
\eea

so that
\bea
m_s^2 \simeq m_{\sigma}^2 =-\lambda_{mix}v^2
\eea

resulting in a very small mass for $m_s$ when $\lambda_{mix} \ll 1$.

\section{Dark Energy}

If the new scalar is associated with a hidden sector, it is natural to assume that its mixings with the visible sector are suppressed by GUT or Planck scale mixings, so we assume a suppression based on a ratio of masses,
\bea
\lambda_{mix}  \approx  - \frac{(m_H)^2}{(M_X)^2} 
\eea
where, $M_X$ is some large mass scale.  Then at the minimum, assuming no vacuum energy contribution from the standard model sector

\bea
|V| \approx  \frac{(m_H)^2}{4(M_{X})^2} v^2 v_{H}^2 - \frac{\lambda_H}{4}v_{H}^4 \approx \frac{(m_H)^4}{16\lambda_H( M_{X})^4} v^4.
\eea

If we assume this energy scale accounts for the observed scale of dark energy in the universe $\approx 2\times10^{-12} GeV$  then we find the appropriate mass scale, $M^*$, is given by, 

\bea
M^* \approx 7.7 \times10^{15} \left(\frac{1}{\lambda_H}\right)^{1/4}  GeV
\eea

Depending upon the value of $\lambda_H$, a GUT scale suppression in mixing for this new scalar will naturally result in a contribution to vacuum energy corresponding to the scale of the dark energy density in the universe.

As we have noted, this vacuum energy contribution from the hidden sector will be negative.  Therefore for this contribution to correspond to the observed dark energy we would need to reside in a false vacuum state, and we need to make a standard assumption that the cosmological constant problem is solved by somehow requiring that the cosmological constant in the true ground state is zero.

The possibility that we reside in a false vacuum state is not unreasonable, given the parameters of this model.  The relevant mass of the new scalar is comparable to the temperature today if $\lambda_H$ is of order $10^{-4}$.  In this case, depending upon the value of $\lambda_H$, thermal effects might preserve a VEV at or near zero for the $s$ field until the universe cools below its current value.  The details of a possible departure from zero over recent cosmological time would then depend on the detailed nature of the hidden sector potential, with possibly observable cosmological implications.

\section{Phenomenology}

The introduction of new nearly massless scalars mixed with the standard model necessarily induces possibly terrestrial experimentally observable effects, from Higgs disappearance phenomena to long range forces \cite{Carroll:1998zi,Dvali:2001dd}.  Given the small mixing envisaged here impacts on Higgs production and decay, and radiative effects at the weak scale will clearly be unobservable.  

Long range forces are another matter however.  Given the weakness of gravity, new nearly massless scalars that might impact upon, say, Weak Equivalence Principle (WEP) measurements, are severely constrained.  Besides direct WEP violation, such scalars will produce forces that depart from $ 1/r^2$ on scales larger than the particles Compton wavelength.  Given the mixing envisaged above to reproduce the dark energy density today, we find the effective long-range force induced by this new scalar in interactions with quarks, for example, introduces a force $F_H$ on scales shorter than the inverse mass of the new scale, compared to  the gravitational force $F_G$ of 

\bea
F_H \simeq \left(\frac{M_{Pl}}{M^*}\right)^2\left(\frac{m_H}{M^*}\right)^4  \frac{1}{\lambda_H} F_G.
\eea

Unfortunately this is well below the scale of forces that can be probed by small $1/r^2$ or weak equivalence principle probes \cite{Adelberger:2003zx,Geraci:2008hb,Sushkov:2011zz,Yang:2012zzb,Wagner:2012}.   

Thus, it would seem at present that the only probe of such new physics involves cosmology.  The dark energy equation of state would depend on whether we continue to reside in a false vacuum (i.e. $w=-1$) or have begun to move away from it.

\section{Model Building and Naturality}

The potential in Eq (1) implies that the new scalar in question is precisely massless before couplings to the Higgs are incorporated.  It is hard to know if this is a reasonable assumption or whether issues of radiative corrections and vacuum stability will be problems without knowing the physics of the hidden sector.  Perhaps scaling or conformal symmetry might preserve this, broken only by couplings to the standard sector.  Imposing a zero mass idoes seem less arbitrary at this point than imposing an arbitrarily fine tuned small mass in the a scalar field Lagrangian in order to generate a viable dark energy scale.  

The possibility of an additional scalar in a hidden sector which gets a small mass by symmetry breaking also introduces a host of new phenomenological possibilities which can be further explored.  For example, if it is charged under a gauge symmetry then there will be associated with it a gauge field with a small mass, which might also mediate possible short range forces.  Alternatively, in the absence of an additional gauge group there could exist a massless Goldstone mode, which might have its own cosmological impacts (i.e. \cite{Weinberg:2013kea,Jeong:2013eza,Anchordoqui:2013pta}).

\section{Conclusions}

The puzzle of dark energy appears, within the context of the standard model, to be the biggest fine-tuning puzzle in physics.  As we have shown here,  it is possible to generate a contribution to the vacuum energy density on the order of the observed dark energy without introducing absurdly low energy scales directly into models, allowing mixing with a hidden sector at a reasonable GUT scale in analogy to a neutrino see-saw. A hierarchy of scales is naturally generated in which the EW scale is the large energy scale, and the dark energy scale is the low energy scale. In this case, we are required to be in a false vacuum today.  Whether our mechanism can be incorporated naturally into a fully phenomenological particle physics framework remains to be seen.  However, it does open the possibility that a solution to at least the fine-tuning puzzle in the context of a solution to the deeper cosmological constant puzzle may involve solving a potentially simpler Higgs-saw puzzle.

\section{Acknowledgements}

We would like to thank Steve Weinberg and Frank Wilczek for helpful comments.  LMK would like to thank the DOE for support as well as ANU for hospitality while this work was being completed.  JBD would like to thank the Louisiana Board of Regents and NSF for support.

\bibliography{Seesawref}

\end{document}